# Leveraging Gaussian Process and Voting-Empowered Many-Objective Evaluation for Fault Identification

Pei Cao, *Student Member, IEEE,* Qi Shuai, and J. Tang, *Member, IEEE*

*Abstract*—Using piezoelectric impedance/admittance sensing for structural health monitoring is promising, owing to the simplicity in circuitry design as well as the high-frequency interrogation capability. The actual identification of fault location and severity using impedance/admittance measurements, nevertheless, remains to be an extremely challenging task. A first-principle based structural model using finite element discretization requires high dimensionality to characterize the high-frequency response. As such, direct inversion using the sensitivity matrix usually yields an under-determined problem. Alternatively, the identification problem may be cast into an optimization framework in which fault parameters are identified through repeated forward finite element analysis which however is oftentimes computationally prohibitive. This paper presents an efficient data-assisted optimization approach for fault identification without using finite element model iteratively. We formulate a many-objective optimization problem to identify fault parameters, where response surfaces of impedance measurements are constructed through Gaussian process-based calibration. To balance between solution diversity and convergence, an $\varepsilon$-dominance enabled many-objective simulated annealing algorithm is established. As multiple solutions are expected, a voting score calculation procedure is developed to further identify those solutions that yield better implications regarding structural health condition. The effectiveness of the proposed approach is demonstrated by systematic numerical and experimental case studies.

This research is supported by the National Science Foundation under Grant CMMI – 1544707.
Pei Cao is with Department of Mechanical Engineering, University of Connecticut, 191 Auditorium Road, Unit 3139, Storrs, CT 06269, USA (email: pei.cao@uconn.edu).
Qi Shuai is with School of Automotive Engineering, Chongqing University, Chongqing, China (e-mail: qishuai@cqu.edu.cn).
Jiong Tang is with Department of Mechanical Engineering, University of Connecticut, 191 Auditorium Road, Unit 3139, Storrs, CT 06269, USA (email: jiong.tang@uconn.edu).

## I. INTRODUCTION

The timely and accurate identification of faults in aerospace, mechanical, marine, and infrastructure systems has received significant recent attention. Different from traditional, offline non-destructive testing and evaluation (NDT&E) techniques, e.g., X-ray inspection, where the effectiveness is limited to the close vicinity of the sensors employed, online structural health monitoring is often facilitated through actuating and then sensing/measuring dynamic responses such as waves/vibrations that can propagate quite far (Farrar and Worden, 2006). This yields much larger coverage area and higher inspection efficiency. The advent of many new transducer materials/devices and the advancement in microelectronics have resulted in rapid progresses in this area. On the other hand, bottlenecks and unique challenges exist. Structures are continuous media, and parameters characterizing structural faults, i.e., location and severity, are continuous variables as well. Hence, structural faults have infinitely many possible patterns/profiles with typically small characteristic lengths, which are further compounded by various uncertainties. Intuitively, the dynamic response data collected by the monitoring system must be in high-frequency range (i.e., with small wavelengths) so features of small-sized faults can be captured. The key issues thus are: 1) how to effectively generate high-frequency sensing data; and 2) how to efficiently and accurately identify fault location and severity from the data (Zhang et al, 2017).

Owing to their two-way electro-mechanical coupling, piezoelectric transducers are commonly used in structural health monitoring (Wang and Tang, 2008; Gao et al, 2018). One class of methods is ultrasonic propagating wave-based, where these transducers are used as actuators and sensors. The change of transient wave (e.g., Lamb wave) patterns, as waves propagate through fault site, can be used to infer fault occurrence. While these methods lead to high detection sensitivity due to the high-frequency nature, it is difficult to use transient responses to identify fault, especially to quantify the severity. The piezoelectric transducers have also been employed in the electrical impedance- or admittance-based methods where a piezoelectric transducer that is integrated (bonded/embedded) with the structure being monitored (Kim and Wang, 2014). In these methods, the piezoelectric transducer is driven by a sinusoidal voltage sweep over a certain frequency range, and the electrical response (i.e., the resulted current) is measured to extract the impedance/admittance information. Owing to the electro-mechanical coupling, the piezoelectric impedance/admittance is directly related to the mechanical impedance of the underlying structure. Thus, the change of piezoelectric impedance/admittance signature with respect to that under the healthy baseline state can be used as fault indicator. These methods have shown effectiveness for a variety of structural faults including crack, corrosion, debonding, joint degradation, etc (Park et al, 2008; Zhou and Zuo, 2012). The impedance/admittance can be measured in high-frequency range. A significant advantage is that in these methods the piezoelectric transducer serves as actuator and sensor simultaneously and the circuitry design is extremely simple requiring essentially only a small resistor, which leads to implementation convenience.

A major hurdle remains. In theory, identifying directly the fault location and severity from stationary responses such as impedances/admittances is possible, as long as a credible first-principle model such as finite element model of the healthy baseline is available. A linearized sensitivity matrix can be derived that links the structural property changes to the changes of harmonic response magnitudes measured. In reality, such an inverse problem is usually severely under-determined. In order to characterize high-frequency impedance/admittance responses accurately, the finite element model must have high dimensionality. To pinpoint fault condition, we often divide the structure into a number of segments where the structural property in each segment is an unknown to be solved (because each segment is susceptible of fault occurrence). Therefore, the model has high dimensionality with a large number of unknowns. Meanwhile, structural faults manifest themselves in structural resonances and anti-resonances. As such, the effective measurements of impedance/admittance changes are limited (Shuai et al, 2017). One potential way to avoid the direct inversion is to convert the identification problem into an optimization problem, where possible property changes in all segments are treated as design parameters. These parameters are updated by minimizing the discrepancy between sensor measurements and model predictions through various optimization techniques in which only forward analyses are performed (Begambre and Laier, 2009; Perera et al, 2010; Cao et al, 2018a and 2018b). The necessary computational cost, however, could be very high. The forward optimization generally requires large number of iterations to converge, while a single run of high-dimensional finite element analysis can be very costly already.

Dynamic response calibration, as a faster alternative to exhaustive finite element analysis, has shown promising aspects in alleviating computational burden by emulating the full-scale finite element model responses. Traditional response surface methods applied for model updating use explicit functions to represent the relation between inputs and outputs. Least square-based techniques are then devised to refine parameters in the polynomial representation (Ren and Chen, 2010; Fang and Perera, 2011; Li and Law, 2011; Chakrarborty and Sen, 2014). More recently, Gaussian process, also referred to as Kriging (Kennedy and O'Hagan, 2001; Rasmussen and Williams, 2006), has gained popularity due to its capability to simulate complicated process subjected to uncertainties. A Gaussian process model is not restricted to certain polynomial form thus allows highly flexible modeling in input-output relation based on statistical expectations and variances over functions. Gao et al (2013) used a Kriging surrogate model to calibrate frequency responses for crack tip location identification in cantilever plates. Yang et al (2017) proposed a similar calibration approach in frequency domain to detect the location and severity of fault in small structures. Wan and Ren (2015) suggested a residual-based Gaussian process model to characterize the relation between residual and updated parameters in frequency domain for finite element model updating. Jin and Jung (2016) formulated a sequential surrogate modeling scheme that constructs multiple response surfaces for finite element model updating. Balafas et al (2018) presented a Gaussian process model in wavelet domain that can infer damage through hypothesis testing. It is worth noting that all these dynamic response calibration methods are applied to natural frequency measurements. Since in practical situation only lower-order natural frequencies can be realistically measured, the case setups in these studies are relatively simple with low dimensionality and the design parameters are discrete with low dimensionality as well. In comparison, in impedance/admittance sensing, considerably more amount of measurements at many frequency points, can be acquired, and a high-dimensional structure is to be identified.

From the underlying physics standpoint, impedance/admittance sensing offers a new opportunity to identify fault parameters more accurately for more complex structures. While the response calibration technique appears to be promising in possibly avoiding iterative finite element analyses in an optimization framework, new issue arises. Although fault effects are reflected in impedance/admittance change at each frequency point theoretically, the actual impedance/admittance measurements respond to a fault condition differently at different frequencies. Therefore, in order to correctly identify fault conditions, one would need to examine the impedance/admittance changes at many frequency points. In other words, in order to take full advantage of the high-frequency impedance/admittance sensing, we need to formulate and then solve efficiently an optimization problem to match response predictions with measurements at many frequency points. It should be noted that many-objective global optimization usually features more than three objectives, while multi-objective optimization refers to that with no more than three objectives. Although it would appear to be easier to resort to weighted summation to solve a single objective optimization (Gao et al, 2013; Wan and Ren, 2015; Yang et al, 2017), weighting selection among objectives is ad-hoc, and the result could easily converge to a meaningless outcome due to multiple local optima, measurement noise and uncertainties.

In this research, we develop a new methodology of fault identification using piezoelectric impedance/admittance sensing. To thoroughly elucidate the health status, a many-objective optimization is formulated to match parametric prediction with measurements at all frequency points of interest. Gaussian process regression is incorporated to construct the response surfaces, which not only significantly reduces computational cost but also yields continuous searching of fault parameters. Our goal in optimization is to find many solutions (owing to the under-determined nature of the problem) that are all optimal. In order to balance between solution convergence and diversity, we establish an $\varepsilon$-dominance enabled many-objective simulated annealing algorithm. Subsequently, inspired by concepts in social statistics, i.e., voting power and majority voting (Taylor and Pacelli, 2008), a voting score calculation framework is employed to evaluate quality of the solutions obtained. As a combination of many-objective optimization and voting score calculation, our proposed many-objective evaluation approach is able to distinguish the solutions that could accurately indicate the health condition of the structure and ultimately provide guidance for further examination.

The rest of this paper is organized as follows. In Section II, we establish the many-objective optimization formulation assisted by Gaussian process regression for piezoelectric impedance active sensing, where the $\varepsilon$-dominance enabled many-objective approach and the voting score calculation are presented in detail. In Section III, the proposed method is evaluated through numerical case studies. Experimental validations are conducted in Section IV. Finally, concluding remarks are given in Section V.

## II. APPROACH FORMULATION

### A. Piezoelectric impedance/admittance active sensing

In piezoelectric impedance/admittance-based fault identification, a piezoelectric transducer circuit is attached to or embedded in a host structure. Harmonic excitation voltage, with frequency referred to as the excitation frequency or driving frequency, is supplied to actuate structural oscillation. The local structural oscillation in turn induces electrical response of the transducer due to electro-mechanical coupling. We can write the equations of motion of the coupled system in the finite element form as (Wang and Tang, 2010),

$$\mathbf{M}\ddot{\mathbf{q}} + \mathbf{C}\dot{\mathbf{q}} + \mathbf{K}\mathbf{q} + \mathbf{K}_{12}Q = \mathbf{0} \tag{1a}$$

$$K_c Q + \mathbf{K}_{12}^T \mathbf{q} = V_{in} \tag{1b}$$

where $\mathbf{M}$, $\mathbf{K}$ and $\mathbf{C}$ are the mass matrix, stiffness matrix and damping matrix, respectively, $\mathbf{q}$ is the structural displacement vector, $\mathbf{K}_{12}$ is the electro-mechanical coupling vector due to piezoelectric effect, $K_c$ is the reciprocal of the capacitance of the piezoelectric transducer, $Q$ is the electrical charge on the surface of the piezoelectric transducer, and $V_{in}$ is the excitation voltage. Clearly in Equation (1), the impedance/admittance of the transducer is directly related to the impedance of the underlying structure and thus can be used as damage indicator. Under harmonic excitation, Equation (1) can be expressed in frequency domain. The admittance (reciprocal of impedance) of the piezoelectric transducer is then given as,

$$Y(\omega) = \frac{\dot{Q}}{V_{in}} = \frac{\omega i}{K_c - \mathbf{K}_{12}^T(\mathbf{K} - \mathbf{M}\omega^2 + \mathbf{C}\omega i)^{-1}\mathbf{K}_{12}} \tag{2}$$

where $\omega$ is the excitation frequency and $i$ is the imaginary unit. In discretized model-based fault identification, structural fault or damage is frequently assumed as local property change, e.g., local stiffness loss. We divide the host structure into $n$ segments and use $\mathbf{k}_{hj}$ to represent the stiffness matrix of the $j$-th segment under healthy condition. The stiffness matrix of the structure when fault occurs can be written as,

$$\mathbf{K}_d = \sum_{j=1}^{n} \mathbf{k}_{hj}(1 - \alpha_j) \tag{3}$$

where $\alpha_j \in [0, 1]$ is the fault index indicating the ratio of stiffness loss in the $j$-th segment. For example, if the $j$-th segment suffers from damage that leads to a 10% stiffness loss, then $\alpha_j = 0.1$, otherwise $\alpha_j = 0$. $\boldsymbol{\alpha} = [\alpha_1, \cdots, \alpha_n]^T$ is the fault index vector. As the piezoelectric transducer and the underlying structure form a coupled system, structural fault will be reflected by the admittance of the piezoelectric transducer,

$$Y_d(\omega, \boldsymbol{\alpha}) = \frac{\dot{Q}_d}{V_{in}} = \frac{\omega i}{K_c - \mathbf{K}_{12}^T(\mathbf{K}_d - \mathbf{M}\omega^2 + \mathbf{C}\omega i)^{-1}\mathbf{K}_{12}} \tag{4}$$

The measured admittance of the structure with fault can then be compared with the baseline admittance to elucidate the health condition. The change of admittance before and after fault occurrence can be written as a function of excitation frequency $\omega$ and damage index vector $\boldsymbol{\alpha}$,

$$\Delta Y(\omega, \boldsymbol{\alpha}) = Y_d(\omega, \boldsymbol{\alpha}) - Y(\omega, \boldsymbol{\alpha} = \mathbf{0})$$
$$= \frac{\omega i (\mathbf{k}_h \mathbf{1} - \mathbf{M}\omega^2 + \mathbf{C}\omega i)(\mathbf{k}_h \boldsymbol{\alpha} - \mathbf{M}\omega^2 + \mathbf{C}\omega i)}{-\mathbf{K}_{12}^T \mathbf{k}_h} \tag{5}$$

In Equation (5), $\mathbf{k}_h = [\mathbf{k}_{h1}, \cdots, \mathbf{k}_{hn}]$, which represents the stiffness sub-matrices of $n$ segments when the structure is healthy. In impedance/admittance-based fault identification, as harmonic voltage excitation is supplied for active sensing, Equation (5) is used iteratively giving different read of $\Delta Y(\omega, \boldsymbol{\alpha})$ when the excitation frequency is swept within certain ranges that cover a number of structural resonances around which physical measurements are taken. In order to characterize high-frequency responses, the finite element model must have high dimensionality.

### B. Data-Assisted Impedance Response Calibration

As indicated in Introduction, direct inverse analysis based on Equation (5) generally yields a severely under-determined problem. One possible solution is to perform repeated forward finite element analyses in the parametric space within an optimization framework to identify fault parameters. In order to render such a procedure computationally tractable, in this sub-section we present a data-assisted meta-modelling approach through Gaussian process (GP) regression (Kennedy and O'Hagan, 2001; Rasmussen and Williams, 2006; Stein, 2012). Essentially, we aim at rapidly constructing the response surfaces in the parametric space through emulations using experimental and/or numerical simulation data.

Gaussian process (GP) regression is an interpolation approach by which various spatial and temporal problems can be modeled (Xia and Tang, 2013; Wang et al, 2017). For impedance-based active sensing, the observed output can be symbolized and denoted as $\Delta Y(\mathbf{x}) = f(\mathbf{x}) + \varphi$, where $f(\mathbf{x})$ is the output of the numerical model, $\mathbf{x}$ is the input vector, and $\varphi$ is the model discrepancy. The additive error $\varphi$ is assumed to follow an independent, identically distributed Gaussian distribution $\varphi \sim N(0, \sigma_n^2)$. A function $\phi(\mathbf{x})$ can be introduced to map the input $\mathbf{x}$ to $f(\mathbf{x})$ as,

$$f(\mathbf{x}) = \phi(\mathbf{x})^T \mathbf{w} \qquad (6)$$

where $\mathbf{w}$ is a vector of unknown parameters. The probability density of the set of training samples $(\Delta \mathbf{Y}, \mathbf{X})$ given $\mathbf{w}$ can then be obtained,

$$p(\Delta \mathbf{Y}|\mathbf{X}, \mathbf{w}) = \prod_{i=1}^{n} p(\Delta Y_i | \mathbf{x}_i, \mathbf{w}) =$$
$$\prod_{i=1}^{n} \frac{1}{\sqrt{2\pi}\sigma_n} \exp(-\frac{(\Delta Y_i - \phi(\mathbf{x}_i)^T \mathbf{w})^2}{2\sigma_n^2}) \sim N(\phi(\mathbf{x})^T \mathbf{w}, \sigma_n^2 \mathbf{I}) \qquad (7)$$

The training samples can be acquired either experimentally or from a credible finite element model. Now we assume a multivariate Gaussian prior over the parameters $\mathbf{w} \sim N(\mathbf{0}, \Sigma_p)$ with zero mean and certain covariance. We can obtain the posterior probability density of $\mathbf{w}$ through Bayes' theorem,

$$p(\mathbf{w}|\mathbf{X}, \Delta \mathbf{Y}) = \frac{p(\Delta \mathbf{Y}|\mathbf{X}, \mathbf{w})p(\mathbf{w})}{p(\Delta \mathbf{Y}|\mathbf{X})} \sim (\sigma_n^{-2} A^{-1} \phi(\mathbf{x}) \Delta \mathbf{Y}, A^{-1}) \qquad (8)$$

where $A = \sigma_n^{-2} \phi(\mathbf{x}) \phi(\mathbf{x})^T + \Sigma_p^{-1}$. Finally, by averaging over all possible parameter values, the predictive distribution of $f_*$ given a test input vector $\mathbf{x}_*$ also follows Gaussian distribution,

$$p(f_* | \mathbf{x}_*, \mathbf{X}, \Delta \mathbf{Y}) = \int p(f_* | \mathbf{x}_*, \mathbf{w}) p(\mathbf{w} | \mathbf{X}, \Delta \mathbf{Y}) d\mathbf{w} \sim$$
$$N(\sigma_n^{-2} \phi(\mathbf{x}_*)^T A^{-1} \phi(\mathbf{x}) \Delta \mathbf{Y}, \phi(\mathbf{x}_*)^T A^{-1} \phi(\mathbf{x}_*)) \qquad (9)$$

Therefore, any finite number of outputs $f_*$ given multiple test inputs $\mathbf{x}_*$ have a joint Gaussian distribution. To define such distribution over the stochastic process $f(\mathbf{x})$, a Gaussian process regression model is developed,

$$f(\mathbf{x}) \sim GP(m(\mathbf{x}), k(\mathbf{x}, \mathbf{x}')) \qquad (10)$$

Equation (10) is fully specified by its mean function $m(\mathbf{x})$ and covariance function or kernel $k(\mathbf{x}, \mathbf{x}')$ where $\mathbf{x}$ and $\mathbf{x}'$ are in either the training or the test sets. For prior $\mathbf{w} \sim N(\mathbf{0}, \Sigma_p)$, the mean and covariance functions that determine the smoothness and variability are written as,

$$m(\mathbf{x}) = E[f] = \phi(\mathbf{x})^T E[\mathbf{w}] = \mathbf{0} \qquad (11)$$

$$k(\mathbf{x}, \mathbf{x}') = E[(f - m)(f' - m')] = \phi(\mathbf{x})^T \Sigma_p \phi(\mathbf{x}') \qquad (12)$$

The joint distribution of observation $\Delta \mathbf{Y}$ and unknown output $\mathbf{f}_*$ given training input set $\mathbf{X}$ and test input set $\mathbf{X}_*$ is then,

$$\begin{bmatrix} \Delta \mathbf{Y} \\ \mathbf{f}_* \end{bmatrix} \sim N\left(\mathbf{0}, \begin{bmatrix} K(\mathbf{X}, \mathbf{X}) + \sigma_n^2 \mathbf{I} & K(\mathbf{X}, \mathbf{X}_*) \\ K(\mathbf{X}_*, \mathbf{X}) & K(\mathbf{X}_*, \mathbf{X}_*) \end{bmatrix}\right) \qquad (13)$$

$K(\mathbf{X}, \mathbf{X}_*)$ denotes the $n \times n_*$ matrix of kernels evaluated at all pairs of training and test points through $k(\mathbf{x}, \mathbf{x}_*)$ for $n$ training samples and $n_*$ test inputs. Finally, the key predictive distribution of Gaussian process regression, i.e., the conditional distribution of $\mathbf{f}_*$, is expressed as

$$p(\mathbf{f}_* | \mathbf{X}_*, \mathbf{X}, \Delta \mathbf{Y}) \sim N(K(\mathbf{X}_*, \mathbf{X})[K(\mathbf{X}, \mathbf{X}) + \sigma_n^2 \mathbf{I}]^{-1} \Delta \mathbf{Y},$$
$$K(\mathbf{X}_*, \mathbf{X}_*) - K(\mathbf{X}_*, \mathbf{X})[K(\mathbf{X}, \mathbf{X}) + \sigma_n^2 \mathbf{I}]^{-1} K(\mathbf{X}, \mathbf{X}_*)) \qquad (14)$$

which is the function-space view of Equation (9). In this research, the input vector is given as $\mathbf{x} = [\omega, \mathbf{\alpha}] \Leftrightarrow [\omega, \alpha_L, \alpha_S]$, where $\omega$ is the excitation frequency, and $\mathbf{\alpha}$ is the fault index vector. The vector $\mathbf{\alpha}$ can be further expressed as $[\alpha_L, \alpha_S]$ for single fault cases, where $\alpha_L$ is the location and $\alpha_S$ is the severity. For example, if a structure is divided into 6 segments and the 3rd segment is subjected to 5% damage (5% stiffness loss), then $\mathbf{\alpha} = [0, 0, 0.05, 0, 0, 0]$ or $\mathbf{\alpha} \Leftrightarrow [\alpha_L, \alpha_S] = [3, 0.05]$. For each given $\omega_j$ ($j = 1, 2, ..., l$) where $l$ is the number of frequency points swept during inspection, if $m$ observations or training data in Gaussian process regression can be obtained as $\mathbf{D}_j = \{(\Delta Y_{ji}, \alpha_{Li}, \alpha_{Si}) | i = 1, 2, ..., m\}$, we can then have $l$ calibrations trained by $\{\mathbf{D}_1, \mathbf{D}_2, ..., \mathbf{D}_l\}$ with a Gaussian process regression model $f(\mathbf{\alpha}) \sim GP(\mathbf{0}, k(\mathbf{\alpha}, \mathbf{\alpha}'))$. One of the most widely-adopted kernel functions is the squared exponential function (Rasmussen and Williams, 2006),

$$k(\mathbf{\alpha}, \mathbf{\alpha}') = \theta_1 \exp\left(\frac{|\mathbf{\alpha} - \mathbf{\alpha}'|^2}{\theta_2}\right) \qquad (15)$$

which is efficient toward cases where the training data is of the same type but in different dimensions. For inputs that have more than one type of feature, such as $[\alpha_L, \alpha_S]$ characterizing location and severity that are different in nature, a well-accepted way to build a kernel is to multiply kernels together (Duvenaud, 2014). In this research, we adopt the product of two squared exponential functions as kernel,

$$k(\mathbf{\alpha}, \mathbf{\alpha}') = \theta_1 \exp\left(\frac{|\mathbf{\alpha} - \mathbf{\alpha}'|^2}{\theta_2}\right) \cdot \theta_3 \exp\left(\frac{|\mathbf{\alpha} - \mathbf{\alpha}'|^2}{\theta_4}\right) \qquad (16)$$

The hyper-parameters $\mathbf{\theta}$ used in kernel are trained by maximizing the marginal likelihood $p(\Delta \mathbf{Y} | \mathbf{X})$, or the log marginal likelihood w.r.t. $\mathbf{\theta}$ and $\sigma_n$,

$$\log p(\Delta \mathbf{Y} | \mathbf{\alpha}) = -\frac{1}{2} \Delta \mathbf{Y}^T (K(\mathbf{\alpha}, \mathbf{\alpha}) + \sigma_n^2 \mathbf{I})^{-1} \Delta \mathbf{Y}$$
$$-\frac{1}{2} \log |K(\mathbf{\alpha}, \mathbf{\alpha}) + \sigma_n^2 \mathbf{I}| - \frac{n}{2} \log 2\pi \qquad (17)$$

The parameters are then evaluated using Markov chain Monte

Carlo method (Neal, 2011) in this study.

Compared to single squared exponential kernel (Equation (15)), the product of squared exponential kernels (Equation (16)) can better represent the training samples in impedance-based fault identification. As shown in Figure 1, admittance changes are calibrated using single squared exponential function as kernel and product of squared exponential functions as kernel, respectively, given 270 training data. The calibration surface is the mean value of the predictive distribution acquired using Equation (14).

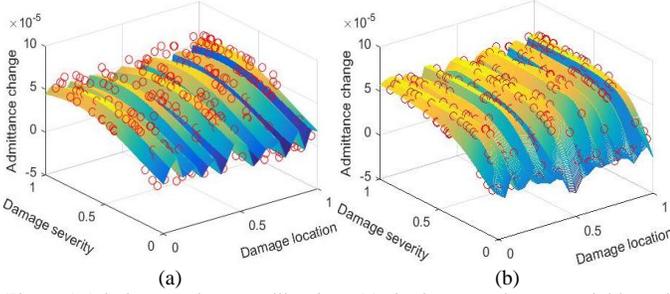

Figure 1 Admittance change calibrations (a) single squared exponential kernel (b) product of squared exponential kernels ( O : training sample)

For $l$ frequencies $\omega_j$ ( $j=1,2,...,l$ ) swept by a piezoelectric transducer in active sensing, if $l$ sets of training data $\mathbf{D}_j$ are available either by experiment or from a finite element model, $l$ calibration surfaces similar to Figure 1(b) can be developed,

$$\begin{aligned} \Delta Y_1^{(c)}(\alpha_L, \alpha_S) | \mathbf{D}_1 \\ \vdots \\ \Delta Y_l^{(c)}(\alpha_L, \alpha_S) | \mathbf{D}_l \end{aligned} \quad (18)$$

where $\Delta Y_j^{(c)}$ represents the output of the reconstructed surface for any input $(\alpha_L, \alpha_S)$ under excitation frequency $\omega_j$.

The proposed method therefore utilizes the regression models to reproduce responses by given different arguments of the response surfaces (health condition of the structure), where the analytical sensitivity matrix to correlate variables with the response is not involved. By minimizing the discrepancy between the predictions made by reconstructed surfaces and the actual measurements, the fault identification problem is essentially cast into an optimization problem. The impedance/admittance changes measured physically under the same $l$ excitation frequencies $\Delta Y_l^{(m)}$ are used to form $l$ objective functions,

$$\begin{aligned} \min J_1 &= \left| \Delta Y_1^{(c)}(\alpha_L, \alpha_S) - \Delta Y_1^{(m)} \right| \\ &\vdots \\ \min J_l &= \left| \Delta Y_l^{(c)}(\alpha_L, \alpha_S) - \Delta Y_l^{(m)} \right| \end{aligned} \quad (19)$$

where $\alpha_S$ and $\alpha_L$ are the design variables of the optimization problem. Consider the case where only one objective function $J_1$ is employed. Minimizing merely $J_1$ will possibly yield a large number of wrong solutions because it is an under-determined problem with only one measurement subjected to error. Clearly, more information regarding the health condition should be taken into consideration by employing more objective functions. This showcases the underlying reason we formulate a many-objective optimization problem. We aim to find the "overlapping consensus" among the available, many objective functions. It is, however, computationally challenging to solve such an optimization problem.

### C. Voting-Empowered Many-Objective Evaluation

Many-objective optimization (MaOO) problems are defined as those with four or more objectives (Deb and Jain, 2014) where the results cannot be directly visualized through graphical means. In comparison, multi-objective optimization problems have two or three objective functions. To illustrate the difficulties associated with solving many-objective optimization problems, we first introduce the Pareto optimality based multi-objective optimization, which has seen extensive research efforts (Zitzler, 1999; Deb et al, 2002; Laumanns et al, 2002; Zhang and Li, 2007; Cao et al, 2017). For multi-objective optimization, the Pareto optimality is defined in a broader sense that no other solution is superior to the Pareto optimal solutions when all objectives are considered. Following this, a general Pareto-based MaOO problem where $n$ objectives are minimized simultaneously is specified as

$$\text{Minimize } \mathbf{f}(\mathbf{x}) = \{f_1(\mathbf{x}),...,f_n(\mathbf{x})\} \quad (20)$$

where $\mathbf{x}$ is the decision vector and $\mathbf{f}$ is the objective vector. When two sets of decision vectors are compared, the concept of dominance is involved. Assuming $\mathbf{a}$ and $\mathbf{b}$ are two decision vectors, the concept of Pareto optimality can be defined as follows: $\mathbf{a}$ dominate $\mathbf{b}$ if:

$$\forall i = \{1,2,...,n\}: f_i(\mathbf{a}) \leq f_i(\mathbf{b}) \quad (21)$$

and

$$\exists j = \{1,2,...,n\}: f_j(\mathbf{a}) < f_j(\mathbf{b}) \quad (22)$$

Any objective function vector, which is neither dominated by any objective function vector in the Pareto optimal set nor dominating any of them, is said to be non-dominated with respect to that Pareto optimal set. The solution that corresponds to the objective function vector is then a member of Pareto optimal set.

In comparison with multi-objective optimization, many-objective evaluation needs to tackle two major additional difficulties (Deb and Jain, 2014; Li et al, 2015):

1) Almost all solutions generated are non-dominated to one another. As the number of objectives increases, even a mediocre solution could be Pareto optimal because it may have small advantages over other solutions in at least one objective, even though the differences are trivial. Consequently, most Pareto optimality-based multi-objective optimization algorithms become inefficient and out of focus when dealing with many objectives. The solution set yielded may be arbitrarily large.

2) It is hard to maintain good diversity among the solution set in high dimensional space. Generally, it is computationally

expensive to evaluate diversity for many objectives. Moreover, the conflict between convergence and diversity is aggravated in high dimension. Therefore, attempts to maintain diversity may hinder the numerical procedure from converging to the optimal solutions.

The difficulties can be alleviated by using a special domination principle that will adaptively discretize the Pareto optimal set and find a well-distributed set of solutions. A good choice to tackle the above-mentioned difficulties is the $\varepsilon$-dominance principle (Laumanns et al, 2002), which alters and discretizes the objective space into boxes defined by the power of $(1+\varepsilon)$,

$$\left\lfloor \frac{\log f_i}{\log(1+\varepsilon)} \right\rfloor \quad (23)$$

Equation (23) projects each objective function vector uniquely to one box, which can neutralize trivial improvements in any objectives. One example is shown in Figure 2, one Pareto optimal solution (1.5, 3.5) in the original objective space is eliminated in the $\varepsilon$-Pareto optimal set because it is merely better in one objective but a lot worse in the other objective compared to solution (1.6, 2.5). And by keeping one solution per box, a bounded size solution set with good diversity could be obtained. The aforementioned difficulties can thus be addressed by the $\varepsilon$-dominance transformation. Accordingly, the dominance relation based on $\varepsilon$-dominance is given in Table 1 where the box operator refers to Equation (23) and $\prec$ is used to denote dominance relation between decision vectors.

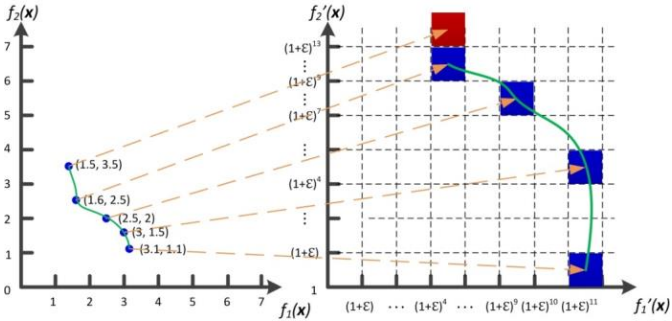

Figure 2 From Pareto optimal to $\varepsilon$-Pareto optimal

Table 1 $\varepsilon$-dominance relations

| Relation | Symbol | Interpretation in $\varepsilon$-objective space |
|---|---|---|
| box(**a**) dominates box(**b**) | $box(\mathbf{a}) \prec box(\mathbf{b})$ | box(**a**) is not worse than box(**b**) in all objectives and better in at least one objective |
| box(**b**) dominates box(**a**) | $box(\mathbf{b}) \prec box(\mathbf{a})$ | box(**b**) is not worse than box(**a**) in all objectives and better in at least one objective |
| Non-$\varepsilon$-dominant to each other | $box(\mathbf{b}) \cong box(\mathbf{a})$ | box(**a**) is worse than box(**b**) in some objectives but better in some other objectives |
| Same box | $box(\mathbf{a}) = box(\mathbf{b})$ | box(**a**) equals box(**b**) |

We incorporate the $\varepsilon$-dominance technique into a previously developed Multi-objective Simulated Annealing algorithm (Cao et al, 2017), hereafter referred to as $\varepsilon$-MOSA/R. The pseudo-code of $\varepsilon$-MOSA/R is provided below.

---

**Algorithm $\varepsilon$-MOSA/R**

Set *Tmax*, *Tmin*, # of iterations per temperature *iter*, cooling rate *α*, *k* = 0
Initialize the *Archive* ($\varepsilon$-Pareto front)
*Current solution* = randomly chosen from *Archive*
**While** (*T* > *Tmin*)
   **For** 1 : *iter*
      Generate a *new solution* vector in the neighborhood of *current solution* vector
      **If** *new solution* falls into the same $\varepsilon$-box as any solutions in the *Archive*
         **If** *new solution* dominates *k* (*k* >= 1) solutions in the *Archive*
            Update
         **Else**
            Action
         **End if**
      **Else if** *new solution* $\varepsilon$-dominates *k* (*k* >= 1) solutions in the *Archive*
         Update
      **Else if** *new solution* $\varepsilon$-dominated by *k* (*k* >= 1) solutions in the *Archive*
         Action
      **Else if** *new solution* and *Archive* are non-$\varepsilon$-dominant to each other
         Update
      **End if**
   **End for**
   *k* = *k*+1
   *T* = (*α*$^k$)\**Tmax*
**End While**

---

**Algorithm Update**

Remove all *k* dominated solutions from the *Archive*
Add *new solution* to the *Archive*
Set *new solution* as *current solution*

---

**Algorithm Action**

**If** *new solution* and *Archive* are non-dominant to each other
   Set *new solution* as *current solution*
**Else**
   **If** *new solution* dominated by *current solution*
     Re-seed
   **Else**
     Simulated Annealing
   **End If**
**End if**

---

**Algorithm Re-seed**

*new solution* is dominated by *k* (*k* >= 1) solutions in the *Archive*
Set *selected solution* as the *i*-th solution $i = \underset{i=1,2,...,k}{\arg\min}(\Delta dom_{i,new})$

**If** $\frac{1}{1+\exp(-\Delta dom_{selected,new} / \max(T,1))}$ > rand(0,1)\*
   Set *selected solution* as *current solution*
**Else**
   **Simulated Annealing**
**End if**
\* rand(0,1) generates a random number between 0 to 1

---

**Algorithm Simulated Annealing**

$$\Delta dom_{avg} = \frac{\sum_{i=1}^{k} \Delta dom_{i,new}}{k}$$

**If** $\frac{1}{1+\exp(\Delta dom_{avg} / T)}$ > rand(0,1)
   Set *new solution* as *current solution*
**End if**

---

In $\varepsilon$-MOSA/R, we use $\varepsilon$-dominance relation as well as the regular dominance relation to compare the new solution, the current solution and Archive. Algorithm Update renews the Archive when a better solution in $\varepsilon$-dominance sense is found and meanwhile assures that only one solution is maintained per $\varepsilon$-box. As Algorithm Re-seed and Algorithm Simulated Annealing are embedded, Algorithm Action takes place when a deteriorated solution is sampled. Instead of abandoning the solution directly, probability relaxations are devised so that either the deteriorated solution is accepted with a certain probability to escape local optima (Simulated Annealing) or the search direction is swerved towards known search space with good solutions for better efficiency (Re-seed). The concept of the amount of domination is used in computing the acceptance probability in Re-seed and Simulated Annealing (Bandyopadhyay et al., 2008). Given two solutions **a** and **b**, the amount of domination is defined as

$$\Delta dom_{\mathbf{a},\mathbf{b}} = \prod_{i=1, f_i(\mathbf{a}) \neq f_i(\mathbf{b})}^{l} (|f_i(\mathbf{a}) - f_i(\mathbf{b})|/R_i) \quad (20)$$

where $l$ is the number of objectives and $R_i$ is the range of the $i$-th objective for normalization. In this research, for all case studies to converge, the total number of iterations of $\varepsilon$-MOSA/R is set as 100,000, $Tmax$ is 100, $Tmin$ is $10^{-4}$, and the cooling rate α is set as 0.8.

Ideally, if the calibration surfaces are perfect without error, using more objectives (i.e., incorporating more measurements) naturally yields solution sets of better accuracy. If sufficient response surfaces are used, the solution set should contain only one solution that matches perfectly the fault scenario. However, owing to modeling and calibration errors, utilizing more objective functions does not necessarily associate with better performance. As seen in Equation (19), $l$ objective functions can be formulated under $l$ excitation frequencies. While $\varepsilon$-MOSA/R introduced could cope with such a many-objective optimization problem, the solution size would increase nonlinearly with $l$ (Duro et al, 2014). Therefore, it could be even harder for a greater number of calibration surfaces to reach the "overlapping consensus" in determining the structural damage. Although using some subsets of available calibrated surfaces can uncover a small set of trustworthy solutions for further analysis, using other subsets may return a large number of erroneous results. Given the difficulties, guiding the algorithm to only a few optimal solutions or making an objective decision becomes critical in many-objective optimization.

In this research, inspired by social statistics (Taylor and Pacelli, 2008), we introduce a novel voting score calculation procedure based on the concepts of majority voting and voting power to evaluate the quality of the solutions generated using different sets of response surfaces as objective functions. As not all those $l$ objective functions are essential or equally important, to reduce variance, $N$ ($N \leq l$) functions are randomly selected from the set as objectives of the many-objective optimization problem denoted as **J**, which can be deemed as input of the many-objective algorithm,

$$\mathbf{A} = \varepsilon\text{-MOSA/R}(\mathbf{J}) \quad (21)$$

where $\mathbf{A} = \{\boldsymbol{\alpha}_a, \boldsymbol{\alpha}_b, \boldsymbol{\alpha}_c, ...\}$ represents the set of Pareto optimal solutions obtained after one many-objective optimization.

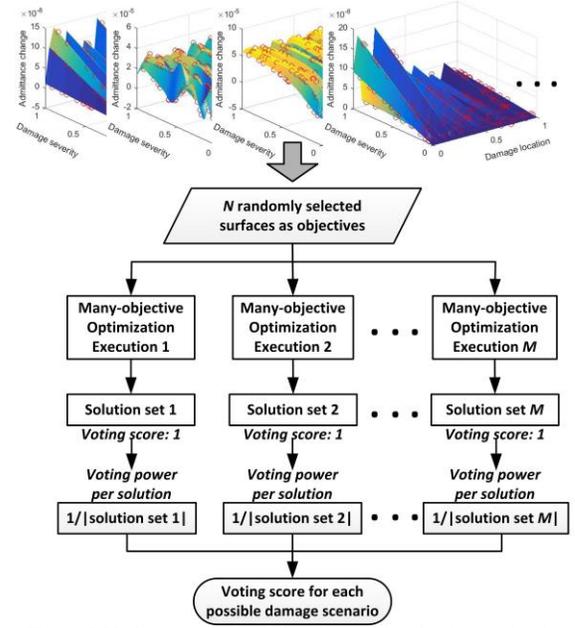

Figure 3 Voting score calculation for multi-objective evaluation

As shown in Figure 3, the many-objective optimization proposed is carried out $M$ times for $N$ randomly selected response surfaces as objective functions different for each execution. Hence, we have,

$$\{\mathbf{A}_1, \mathbf{A}_2, ..., \mathbf{A}_M\} = \varepsilon\text{-MOSA/R}(\{\mathbf{J}_1, \mathbf{J}_2, ..., \mathbf{J}_M\}) \quad (22)$$

We use $\mathbf{A}_i$ to represent the solution set of the $i$-th execution of the optimization given objective function set $\mathbf{J}_i$. The voting score for a specific solution is calculated as,

$$\text{vs}(\boldsymbol{\alpha}_a) = \sum_{i=1}^{M} |\boldsymbol{\alpha}_a \cap \mathbf{A}_i| / |\mathbf{A}_i| \quad (23)$$

where $|\boldsymbol{\alpha}_a \cap \mathbf{A}_i|$ equals to 1 if $\mathbf{A}_i$ contains $\boldsymbol{\alpha}_a$. For example, if $\boldsymbol{\alpha}_a$ appears in optimal solution set $\mathbf{A}_1$, $\mathbf{A}_3$ and $\mathbf{A}_4$, then $\text{vs}(\boldsymbol{\alpha}_a) = 1/|\mathbf{A}_1| + 1/|\mathbf{A}_3| + 1/|\mathbf{A}_4|$. As given in Equation (23), the solution set obtained after each optimization practice is assigned a total voting score of one, meaning that the more solutions there are in one solution set, the less voting power per solution. The rationale behind such design is that we want to grant larger voting power to the solutions in smaller solution sets which are considered to be less affected by error. Thereafter, the scores assigned are added altogether for each possible damage scenario and the ones with higher voting scores are more likely to give accurate implications about the true structural damage. As a result, we look for indications made by the calibrations rather than a decisive result, which is prone to error and not easy to obtain owing the under-determined nature of fault identification systems. Notably, by keeping one less digit after the decimal point in terms of damage severity $\alpha_L$, we are able to investigate the voting scores for possible severity ranges, which could further avert practitioners from investigating inaccurate results.

$$\text{vs}(\hat{\mathbf{\alpha}}_a) = \text{vs}(\alpha_{aL}, \text{round}(\alpha_{aS})) = \sum_{i=1}^{M} |\hat{\mathbf{\alpha}}_a \bigcap \mathbf{A}_i| / |\mathbf{A}_i| \quad (24)$$

where the rounding operator (i.e., 'round' in Equation (24)) approximates a fractional decimal number by one with one less digit, and $|\hat{\mathbf{\alpha}}_a \bigcap \mathbf{A}_i|$ here gives the number of solutions belong to both $\hat{\mathbf{\alpha}}_a$ and $\mathbf{A}_i$.

Recall that voting score calculation is designed to endow those solutions more voting power when the solution set is small. We go one step further by withdrawing the voting scores from the solution sets that exceed the average size of all solution sets instead of equally assigning each solution set a voting score of one.

$$\text{vs}_{partial}(\mathbf{\alpha}_a) = \sum_{i=1}^{M} \frac{\text{I}(|\mathbf{A}_i| \leq |\bar{\mathbf{A}}|) \cdot (|\mathbf{\alpha}_a \bigcap \mathbf{A}_i|)}{|\mathbf{A}_i|} \quad (25)$$

$$\text{vs}_{partial}(\hat{\mathbf{\alpha}}_a) = \sum_{i=1}^{M} \frac{\text{I}(|\mathbf{A}_i| \leq |\bar{\mathbf{A}}|) \cdot (|\hat{\mathbf{\alpha}}_a \bigcap \mathbf{A}_i|)}{|\mathbf{A}_i|} \quad (26)$$

where $\text{I}(|\mathbf{A}_j| \leq |\bar{\mathbf{A}}|)$ is a logic operation that the value of it is 1 if the argument is true and 0 otherwise. By applying either Equation (25) or (26) for post-processing, a higher level of separation between insightful solutions and trivial solutions could be gained.

In this study, we propose four voting score calculation heuristics (Equations (23-26)) that essentially serve as four decision making strategies attempting to identify or isolate possible fault scenarios for further inspection. The randomness introduced when selecting response surfaces (Figure 3) as objective functions has desirable characteristics. It not only makes the evaluation scheme robust to outliers, but also gives useful internal estimates of noise such that we can withdraw voting scores from certain solution sets (Equations (25)(26)). Moreover, it is compatible with parallel computation. Combined with Gaussian process regression and many-objective optimization, the proposed data-assisted many-objective evaluation framework is illustrated in details through numerical and experimental case studies in Section III and Section IV, respectively.

## III. NUMERICAL CASE STUDIES

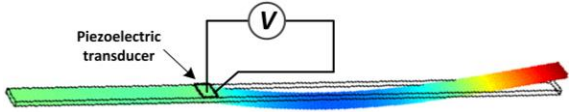

Figure 4 Illustration of structure for numerical case studies

In this section, we carry out two numerical case studies using the proposed methodology to gain insights. The structure of interest is an aluminum cantilevered plate (Figure 4) with the following properties: length $0.561\,\text{m}$, width $0.01905\,\text{m}$, thickness $0.0014\,\text{m}$, density $2700\,\text{kg/m}^3$, and Young's modulus $68.9\,\text{GPa}$. A piezoelectric transducer is attached to the middle left of the plate, i.e., 0.18 m from the fixed end. The properties of the piezoelectric transducer are: length $0.015\,\text{m}$, width $0.01905\,\text{m}$, thickness $0.0014\,\text{m}$, Young's moduli $Y_{11} = 86\,\text{GPa}$ and $Y_{33} = 73\,\text{GPa}$, density $9500\,\text{kg/m}^3$, piezoelectric constant $-1.0288 \times 10^9\,\text{V/m}$, and dielectric constant $\beta_{33} = 1.3832 \times 10^8\,\text{m/F}$. The finite element model of the plate contains 11,250 20-node hexahedron elements, the size of which is smaller than the shortest wavelength of the response involved in this study. The plate is further evenly divided into 25 segments along the lengthwise direction; each is a possible damage location. In structural health monitoring using impedance or admittance measurements, the response changes due to damage occurrence are most evident around the resonant peaks. In the following numerical case studies, we acquire admittance measurements at 40 excitation frequencies around the plate's $14^{th}$, $16^{th}$, $21^{st}$ and $23^{rd}$ natural frequencies. Specifically, the admittance values at 40 evenly distributed excitation frequencies in the ranges 1886.6 Hz to 1890.4 Hz, 2423.7 Hz to 2428.5 Hz, 3694.6 Hz to 3702.0 Hz and 4438.7 Hz to 4447.6 Hz are employed. Identical for each frequency, 270 randomly generated fault scenarios are emulated for the calibration of impedance response surface. The sampling range is specified as 1 to 25 for location and 0 to 0.1 for severity. A set of impedance measurements is produced by sweeping through the pre-specified excitation frequency points numerically. It is worth noting that in actual implementation, we can directly utilize experimentally acquired measurements in lieu of the numerical ones. The data sampled from the numerical model is contaminated by $\pm$ 0.15% standard Gaussian uncertainties to demonstrate the effectiveness of the proposed approach. Here we assume two fault scenarios, i.e., damage occurring at the $13^{th}$ segment with severity $\alpha_{13} = 0.0600$ (6.00% stiffness loss in $13^{th}$ segment), and at $22^{nd}$ segment with severity $\alpha_{22} = 0.0857$ (8.57% stiffness loss), which are randomly selected.

Figure 5 plots all 40 impedance response surfaces reconstructed through Gaussian process regression outlined in Section II-B, which serve as 40 objective functions. The two horizontal axes indicate damage location and severity (normalized), and the vertical axis indicates the admittance change measured by the piezoelectric transducer circuit.

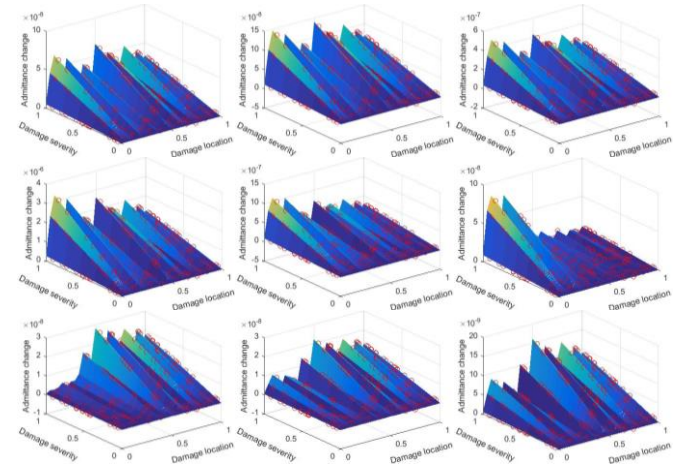

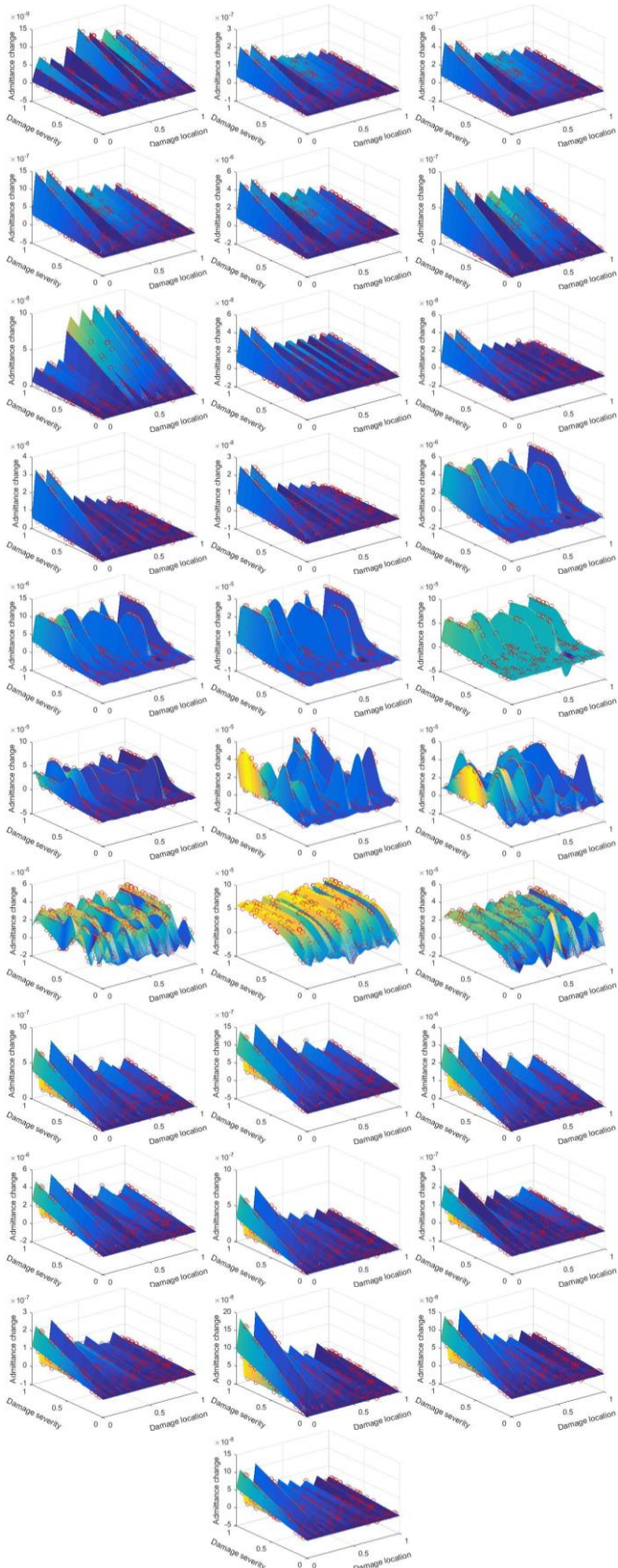

Figure 5 Calibrated response surfaces for 40 excitation frequencies (○ denotes training point).

Based on the many-objective evaluation approach outlined in Section II-C (Figure 3), for each optimization practice, 10 ($N$=10) surfaces out of 40 are randomly selected as objective functions, and the many-objective optimization is executed 30 ($M$=30) times. The parameter $N$ is selected in accordance with the capacity of the many-objective optimization algorithm implemented, and the value of $M$ should be set as large as possible for robustness. In this study, we use $M$=30 for illustration. In other words, a total of 30 voting scores are assigned to possible solutions obtained in 30 many-objective optimization practices.

### A. 6.00% Stiffness Loss in 13$^{th}$ Segment

We first investigate the case where the 13$^{th}$ segment suffers from 6.00% stiffness loss. Here, the post-processing of the MaOO evaluation results introduced in Section II-C warrants detailed discussion. After performing the many-objective optimization, we obtain 369 solutions corresponding to 369 possible damage scenarios. Voting score calculation (Equation (23)) is then carried out successively accrediting score to each solution as quality quantification. As shown in Figure 6, if we consider the solution with the highest voting score as damage identified, it agrees with the actual damage with only 0.0001 discrepancy in stiffness loss ratio.

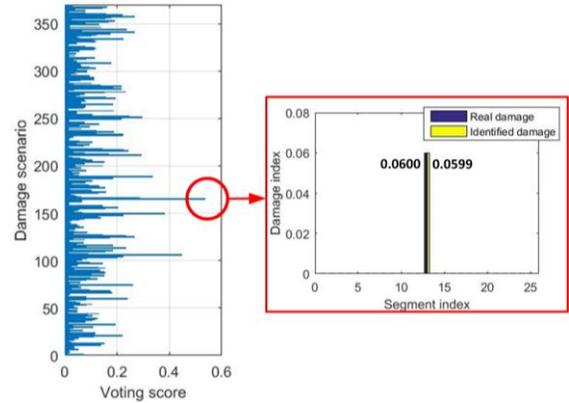

Figure 6 Voting scores for 369 fault scenarios and the one with the highest score.

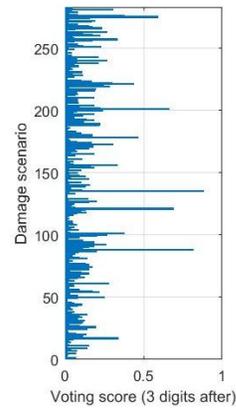

Figure 7 Voting scores for 282 damage severity ranges

In the original voting score calculation, four digits are kept in ratio form in terms of damage severity. Then by keeping one less digit after the decimal point following Equation (24), we are able to investigate the voting scores for possible severity ranges, as illustrated in Figure 7. Out of 282 damage severity ranges, some are more distinct from the others compare to the results shown in Figure 6, which could be considered as candidate solutions. Figure 8 visualizes the severity range with the highest voting score, in which the induced damage is indeed included.

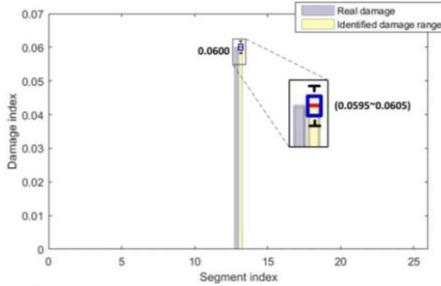

Figure 8 Identified damage range (the severity range with the highest voting score)

We then withdraw the voting scores from the solution sets that exceed the average size of all solution sets as suggested in Equation (25). A total of 16 voting scores are to be distributed among possible solutions found by the many-objective optimization algorithm. As shown in Figure 9(a), some of the fault scenarios are now affiliated with zero voting scores. Comparably, the solutions with higher voting scores are more significantly separated. Similarly, the voting scores for severity ranges can be inspected by grouping certain damage severities together (Equation (26)) and, as illustrated in Figure 9(b), a greater separation is achieved due to the aggregation of voting scores among similar solutions. The purpose of getting higher level of separation hinges on the fact that such fault identification scheme is under-determined due to deficient measurements, uncertainties and errors. Thus, multiple solutions are expected. The proposed data-assisted many-objective evaluation endeavors to isolate a small number of possible solutions by their voting scores for further inspections.

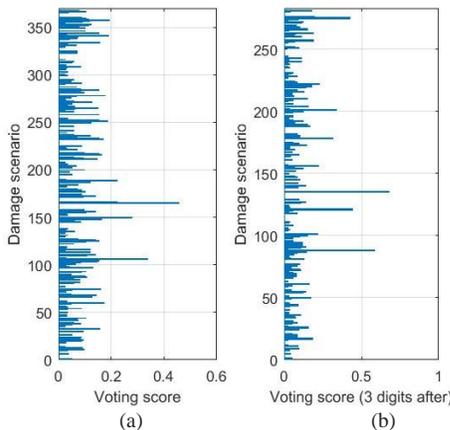

Figure 9 (a) Partial voting scores for 369 damage scenarios (b) Partial voting scores for 282 damage severity ranges

Table 2 lists the fault scenarios with top five highest voting scores calculated following four different heuristics introduced in Section II-C. As shown in the table, the ones with the highest scores all match or cover the true fault scenario. The percentage of voting score out of all voting scores being allotted is also reported in Table 2. For the result that indicates true damage the best, the voting score percentage increases either by grouping severities to severity ranges or assign zero voting scores to large solution sets, meaning a higher level of separation or a higher level of confidence is achieved.

Nevertheless, when prior knowledge is unavailable, solutions with relatively higher scores all should be considered as possible fault scenarios. The proposed voting score scheme filters out most scenarios. Therefore, even though one single certain solution is hard, if not impossible, to obtain, only a few need to be examined with the help of many-objective evaluation, and the one with the highest voting score is more likely to match the true fault scenario.

Table 2 Top five fault scenarios with highest voting scores

| I: Voting score (Equation (23)) | | | |
|---|---|---|---|
| Fault scenario | | Voting score | Score % 30 overall |
| Segment | Severity | | |
| 13 | 0.0599 | 0.5351 | 1.784% |
| 10 | 0.0522 | 0.4474 | 1.491% |
| 12 | 0.0532 | 0.3822 | 1.274% |
| 15 | 0.0505 | 0.3340 | 1.113% |
| 17 | 0.0880 | 0.2922 | 0.974% |
| II: Voting score for severity range (Equation (24)) | | | |
| Fault scenario | | Voting score | Score % 30 overall |
| Segment | Severity range | | |
| 13 | 0.0595~0.0605 | 0.8822 | 2.941% |
| 10 | 0.0515~0.0525 | 0.8137 | 2.712% |
| 12 | 0.0525~0.0535 | 0.6881 | 2.294% |
| 17 | 0.0875~0.0885 | 0.6634 | 2.211% |
| 25 | 0.0575~0.0585 | 0.5896 | 1.965% |
| III: Partial voting score (Equation (25)) | | | |
| Fault scenario | | Voting score | Score % 16 overall |
| Segment | Severity | | |
| 13 | 0.0599 | 0.4579 | 2.862% |
| 10 | 0.0522 | 0.3373 | 2.108% |
| 12 | 0.0532 | 0.2790 | 1.744% |
| 15 | 0.0505 | 0.2239 | 1.399% |
| 13 | 0.0602 | 0.2215 | 1.384% |
| IV: Partial voting score for severity range (Equation (26)) | | | |
| Fault scenario | | Voting score | Score % 16 overall |
| Segment | Severity range | | |
| 13 | 0.0595~0.0605 | 0.6794 | 4.246% |
| 10 | 0.0515~0.0525 | 0.5866 | 3.666% |
| 12 | 0.0525~0.0535 | 0.4434 | 2.771% |
| 25 | 0.0575~0.0585 | 0.4251 | 2.657% |
| 17 | 0.0875~0.0885 | 0.3400 | 2.125% |

### B. 8.57% Stiffness Loss in 22$^{nd}$ Segment

To further demonstrate the proposed approach, we investigate a second numerical case where the 22$^{nd}$ segment suffers from 8.57% stiffness loss. After performing the many-objective optimization, we first obtain 491 possible fault scenarios (Figure 10(a)), which can be grouped into 365 severity ranges following Equation (24) (Figure 10(b)). Aiming at separating the solutions with relatively higher voting scores, the post-processing heuristics given in Equations (25) and (26) are implemented here as shown in Figure 10(c) and 10(d), respectively. The results with top five highest voting scores are ranked in Table 3. Similar to the results reported in Section III-A, in this case study, the ones with the highest scores all agree with or cover the true fault scenario.

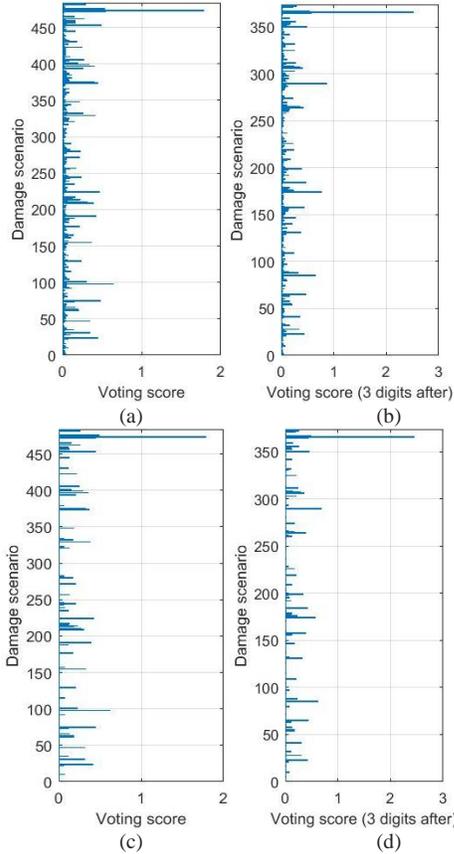

Figure 10 (a) Voting scores for 491 fault scenarios (b) Voting scores for 365 damage severity ranges (c) Partial voting scores for 491 fault scenarios (d) Partial voting scores for 365 damage severity ranges

Table 3 Top five fault scenarios with highest voting scores

| I: Voting score (Equation (23)) | | | |
|---|---|---|---|
| Fault scenario | | Voting score | Score % 30 overall |
| Segment | Severity | | |
| 22 | 0.0856 | 1.0044 | 3.348% |
| 25 | 0.0813 | 0.7574 | 2.525% |
| 25 | 0.0923 | 0.6529 | 2.176% |
| 22 | 0.0843 | 0.4427 | 1.476% |
| 13 | 0.0946 | 0.4389 | 1.463% |
| II: Voting score for severity range (Equation (24)) | | | |
| Fault scenario | | Voting score | Score % 30 overall |
| Segment | Severity range | | |
| 22 | 0.0855~0.0865 | 1.4931 | 4.977% |
| 25 | 0.0915~0.0925 | 0.8741 | 2.914% |
| 25 | 0.0805~0.0815 | 0.7573 | 2.524% |
| 15 | 0.0675~0.0685 | 0.7321 | 2.440% |
| 13 | 0.0875~0.0885 | 0.6866 | 2.289% |
| III: Partial voting score (Equation (25)) | | | |
| Fault scenario | | Voting score | Score % 17 overall |
| Segment | Severity | | |
| 22 | 0.0856 | 1.0044 | 5.908% |
| 25 | 0.0813 | 0.7433 | 4.372% |
| 25 | 0.0923 | 0.5045 | 2.968% |
| 22 | 0.0843 | 0.4027 | 2.369% |
| 22 | 0.0863 | 0.3533 | 2.078% |
| IV: Partial voting score for severity range (Equation (26)) | | | |
| Fault scenario | | Voting score | Score % 17 overall |
| Segment | Severity range | | |
| 22 | 0.0855~0.0865 | 1.444 | 8.494% |
| 25 | 0.0805~0.0815 | 0.7433 | 4.372% |
| 25 | 0.0915~0.0925 | 0.7257 | 4.269% |
| 22 | 0.0835~0.0845 | 0.5675 | 3.338% |
| 13 | 0.0935~0.0945 | 0.5674 | 3.338% |

Here we also investigate the effectiveness of the proposed voting score calculation (Equation (23)) in discriminating possible damage scenarios. The idea of voting has also been used in ensemble learning such as random forest (Breiman, 2001) and pattern recognition (Lam and Suen, 1997) to combine different sets of result where majority voting is implemented. Compared to the proposed voting score strategy, majority voting considers the one damage scenario that appears the most in all solution sets as the indication of true damage. Figure 11 compares the solution with the highest voting score, which concur with the true damage scenario, to the solution that appears the most in all solution sets. As revealed in Table 4, the voting score calculation successfully re-adjusts the weighting among all solutions. The voting score heuristics manage to rank them essentially based on their quality and thus have better performance in terms of identifying the true damage. Indeed, the solution with the highest voting score is not only among the solutions that appear the most in the results of many-objective optimizations, but also is less affected by error trade-offs because it appears mostly in small solution sets which are considered more insightful with less conflicting objective functions. After all, the objective functions should not be contradicted with each other ideally without error.

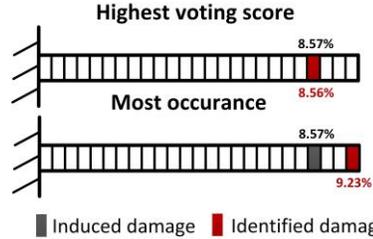

Figure 11 The fault scenario with the highest voting score vs. the damage scenario with the most occurrence

Table 4 Top five fault scenarios: highest voting scores vs. most occurrences

| I: Proposed voting score strategy | | |
|---|---|---|
| Damage scenario | | Voting score 30 overall |
| Segment | Severity | |
| 22 | 0.0856 | 1.0044 |
| 25 | 0.0813 | 0.7574 |
| 25 | 0.0923 | 0.6529 |
| 22 | 0.0843 | 0.4427 |
| 13 | 0.0946 | 0.4389 |
| II: Majority voting (most occurrence) | | |
| Damage scenario | | Occurrence 1289 overall |
| Segment | Severity | |
| 25 | 0.0923 | 17 |
| 15 | 0.0682 | 16 |
| 13 | 0.0946 | 15 |
| 22 | 0.0856 | 14 |
| 13 | 0.0880 | 14 |

## IV. EXPERIMENTAL VALIDATION

In this section, experimental case studies using physical measurements of piezoelectric admittance are carried out. The experimental setup, geometry measures and material parameters are consistent with those used in the numerical analysis in Section III. Figure 12 shows the experimental setup. To obtain the admittance of the piezoelectric circuit, a resistor of $100\,\Omega$ is serially-connected to the transducer to measure the voltage drop, which is further used to extract the current in the circuit. An Agilent 35670A signal analyzer is employed, where the source channel is used to generate sinusoidal voltage sent to piezoelectric transducer denoted as $V_{in}$, and the output voltage across the resistor is recorded as $V_{out}$. Hence, the admittance can be obtained as $Y = I/V_{in} = V_{out}/R_s V_{in}$. In experimental case studies, we acquire measurement samples using 18 excitation frequencies around the plate's $14^{th}$ and $21^{st}$ natural frequencies. That is, 10 evenly distributed frequencies from the range 1886.6 Hz to 1890.4 Hz and 8 evenly distributed frequencies from the range 3696.2 Hz to 3702.0 Hz are acquired.

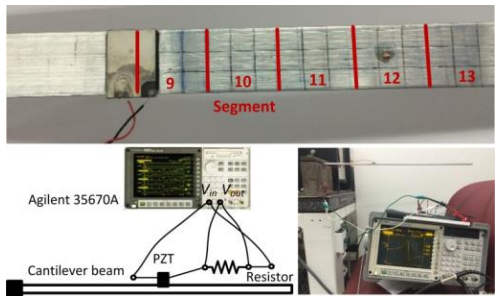

Figure 12 Experiment Setup

Identical for each frequency, 150 randomly generated damage scenarios are emulated for impedance response surface calibration using the corresponding numerical model. Figure 13 illustrates all 18 impedance response surfaces reconstructed by Gaussian process regression. In order to reduce the unwanted variations and uncertainties in this case illustration, instead of disassembling and cutting the plate to reduce the local stiffness, we add small masses to emulate the damage occurrence. Mathematically, adding a small mass can result in the same resonant frequency shift and admittance change as a local stiffness reduction would. In the first experiment, a 0.6 g mass is attached to the $14^{th}$ segment of the plate, which causes admittance change equivalent to a 0.28% local stiffness loss. In the second experiment, the same mass is attached to the $12^{th}$ segment, which is equivalent to a 0.16% local stiffness loss.

Based on the methodology proposed, the many-objective optimization is executed 10 times, and for each optimization execution, 10 surfaces out of 18 are randomly selected as objective functions. In other words, a total of 10 voting scores are assigned to solutions obtained.

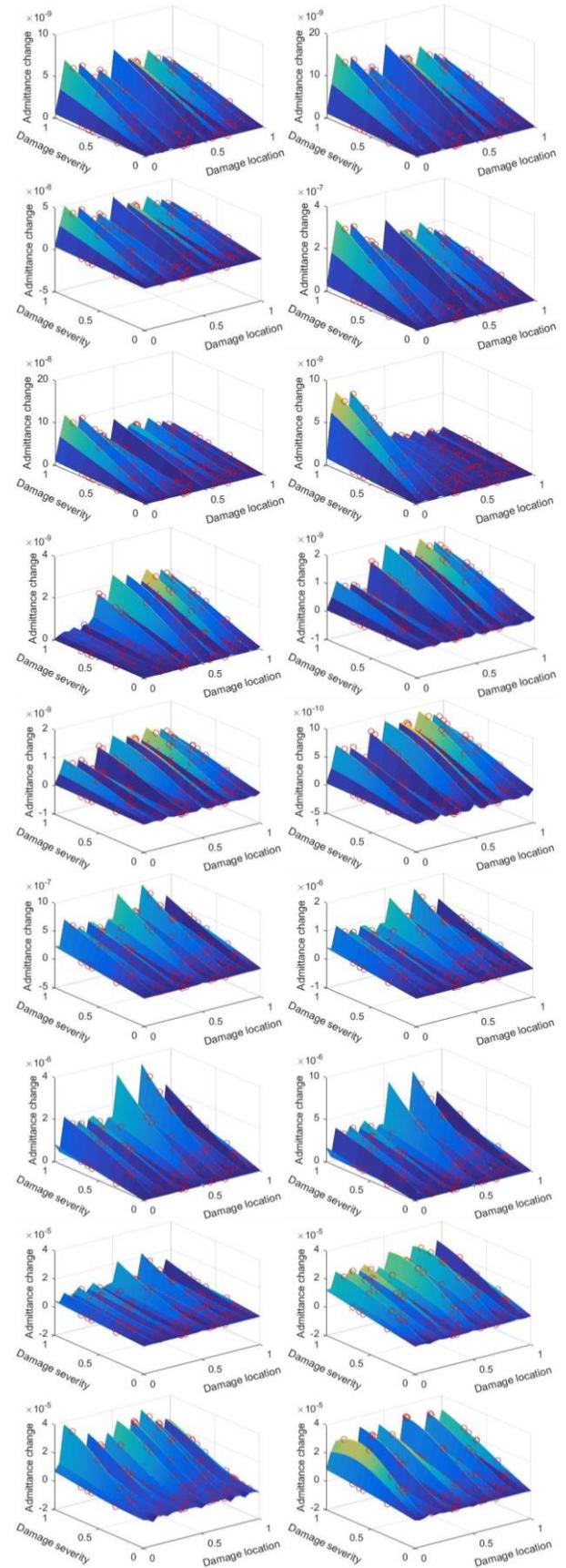

Figure 13 Calibrated response surfaces for 18 excitation frequencies from small to large (○ denotes training point)

## A. 0.28% Stiffness Loss in 14th Segment

We first perform the experimental case study where the 14th segment is subjected to a 0.28% equivalent stiffness loss. Figure 14(a) plots the voting scores (Equation (23)) for all possible fault scenarios after many-objective optimization. As uncertainties such as modeling error and measurement error are present inevitably in experimental case studies, we cannot easily distinguish the better solutions. However, by examining the solutions based on the severity ranges they fall into following the heuristic given in Equation (24) (Figure 14(b)), a few solutions stand out. As shown in Figure 15, if we consider the solutions with the highest voting scores as damage identified or the damage severity range identified, accurate predictions can be achieved. In practice, the solutions with relatively higher voting scores should be considered as candidates. Such candidate set provided by the proposed evaluation approach serves as the starting point for further inspections, which streamlines the typical procedure of inspection and maintenance in engineering practice.

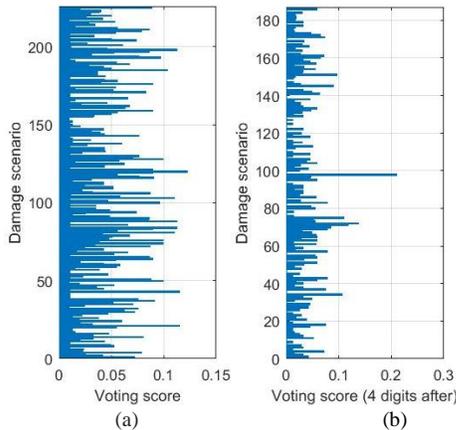

Figure 14 (a) Voting scores for 225 fault scenarios (b) Voting scores for 186 damage severity ranges

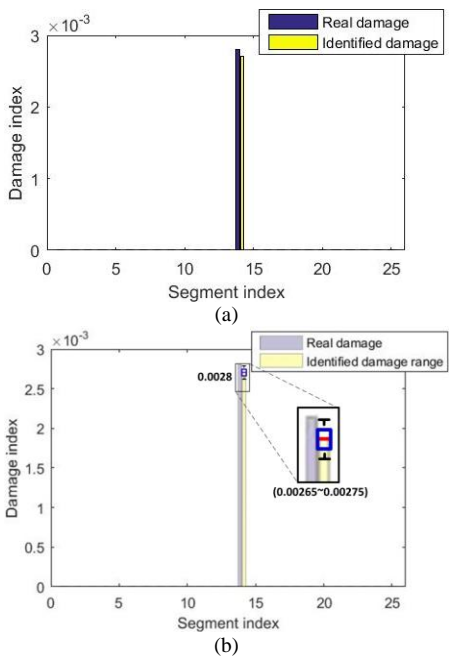

Figure 15 (a) Identified damage (the damage scenario with the highest voting score) (b) Identified damage range (the severity range with the highest voting score)

Next, by assigning zero voting scores to the solution sets that exceed the average size of all solution sets (Equation (25)), we can probe some of the solutions provided by certain solution sets which are considered of better quality. As shown in Figure 16, a higher level of distinction is achieved among solutions. The results are reported in Table 5 where the top five damage scenarios with highest voting scores are demonstrated.

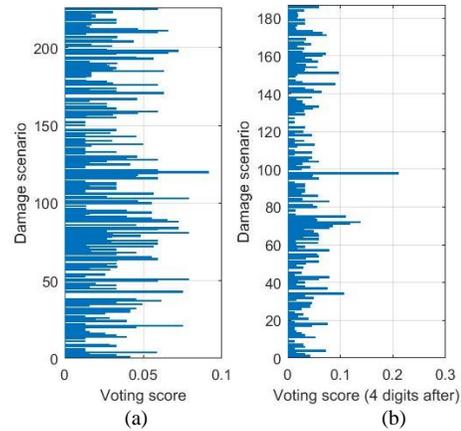

Figure 16 (a) Partial voting scores for 225 fault scenarios (b) Partial voting scores for 186 damage severity ranges.

Table 5 Top five fault scenarios with highest voting scores

| I: Voting score (Equation (23)) | | | |
|---|---|---|---|
| Fault scenario | | Voting score | Score % |
| Segment | Severity | | 10 overall |
| 14 | 0.00271 | 0.1223 | 1.223% |
| 5 | 0.00609 | 0.1153 | 1.153% |
| 3 | 0.00294 | 0.1153 | 1.153% |
| 22 | 0.00492 | 0.1126 | 1.126% |
| 11 | 0.00719 | 0.1126 | 1.126% |
| **II: Voting score for severity range (Equation (24))** | | | |
| Fault scenario | | Voting score | Score % |
| Segment | Severity range | | 10 overall |
| 14 | 0.00265~0.00275 | 0.3016 | 3.016% |
| 11 | 0.00715~0.00725 | 0.2078 | 2.078% |
| 11 | 0.00705~0.00715 | 0.1882 | 1.882% |
| 21 | 0.00915~0.00925 | 0.1777 | 1.777% |
| 5 | 0.00515~0.00525 | 0.1671 | 1.671% |
| **III: Partial voting score (Equation (25))** | | | |
| Fault scenario | | Voting score | Score % |
| Segment | Severity | | 6 overall |
| 14 | 0.00271 | 0.0917 | 1.528% |
| 6 | 0.00458 | 0.0789 | 1.315% |
| 12 | 0.00645 | 0.0787 | 1.312% |
| 11 | 0.00622 | 0.0787 | 1.312% |
| 5 | 0.00609 | 0.0747 | 1.245% |
| **IV: Partial voting score for severity range (Equation (26))** | | | |
| Fault scenario | | Voting score | Score % |
| Segment | Severity range | | 6 overall |
| 14 | 0.00265~0.00275 | 0.2099 | 3.498% |
| 11 | 0.00715~0.00725 | 0.1372 | 2.287% |
| 11 | 0.00705~0.00715 | 0.1176 | 1.960% |
| 11 | 0.00795~0.00805 | 0.1102 | 1.837% |
| 5 | 0.00515~0.00525 | 0.1064 | 1.773% |

## B. 0.16% Stiffness Loss in 12$^{th}$ Segment

The second experimental study concerns the case where the 12$^{th}$ segment is subjected to a 0.16% equivalent stiffness loss. Figure 17 plots the voting scores for all 139 possible fault scenarios and 71 severity ranges after the many-objective evaluation following Equation (23) and (24). It is noticed that a small set of solutions clearly maintains a margin over the rest of the solutions in terms of voting score. As shown in Figure 18, the solution with the highest voting score delivers close indication about the health condition of the structure. And if we consider the severity range with the highest voting score as the identified damage severity range, it also covers the true damage scenario.

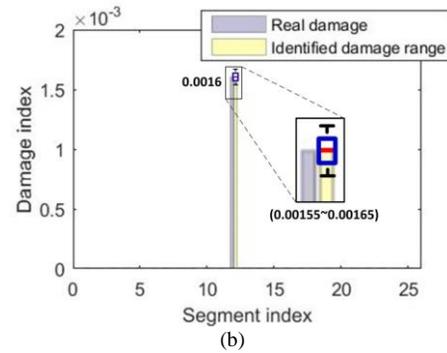

(b)

Figure 18 (a) Identified fault (the fault scenario with the highest voting score) (b) Identified damage range (the severity range with the highest voting score)

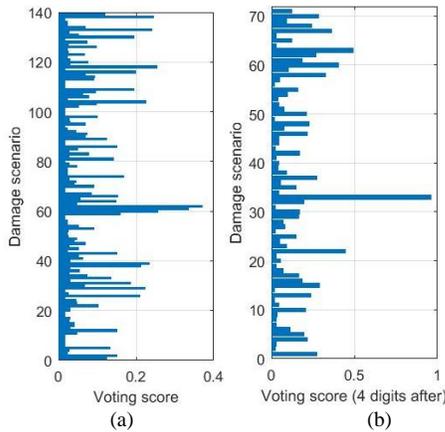

Figure 17 (a) Voting scores for 139 fault scenarios (b) Voting scores for 71 damage severity ranges

Similar to the preceding case studies, zero voting scores are assigned to the solution sets that exceed the average size (Equation (25)), which produces a more polarized voting score distribution shown in Figure 19. As can be seen in Table 6, for all four post-processing means, the ones with the highest voting scores all make accurate implications of the health condition of the structure.

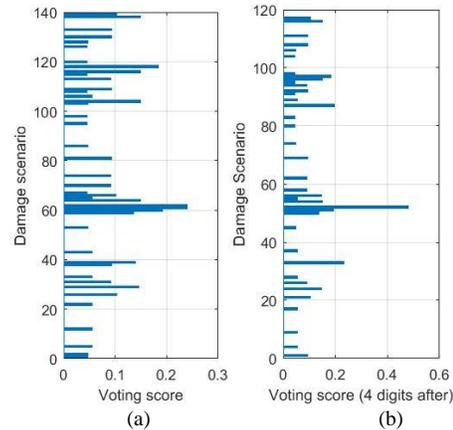

(a)                                         (b)

Figure 19 (a) Partial voting scores for 139 fault scenarios (b) Partial voting scores for 71 damage severity ranges

Table 6 Top five fault scenarios with highest voting scores

| I: Voting score (Equation (23)) | | | |
|---|---|---|---|
| Damage scenario | | Voting score | Score % 10 overall |
| Segment | Severity | | |
| 12 | 0.00161 | 0.3696 | 3.696% |
| 12 | 0.00157 | 0.3355 | 3.355% |
| 12 | 0.00154 | 0.2554 | 2.554% |
| 22 | 0.00161 | 0.2536 | 2.536% |
| 25 | 0.00027 | 0.2446 | 2.446% |
| II: Voting score for severity range (Equation (24)) | | | |
| Damage scenario | | Voting score | Score % 10 overall |
| Segment | Severity range | | |
| 12 | 0.00155~0.00165 | 0.7050 | 7.050% |
| 8 | 0.00135~0.00145 | 0.4467 | 4.467% |
| 21 | 0.00145~0.00155 | 0.3232 | 3.232% |
| 1 | 0.00105~0.00115 | 0.2734 | 2.734% |
| 22 | 0.00155~0.00165 | 0.2705 | 2.705% |
| III: Partial voting score (Equation (25)) | | | |
| Damage scenario | | Voting score | Score % 4 overall |
| Segment | Severity | | |
| 12 | 0.00160 | 0.2395 | 5.988% |
| 12 | 0.00157 | 0.2395 | 5.988% |
| 12 | 0.00154 | 0.1919 | 4.798% |
| 22 | 0.00161 | 0.1840 | 4.600% |
| 25 | 0.00027 | 0.1486 | 3.715% |
| IV: Partial voting score for severity range (Equation (26)) | | | |
| Damage scenario | | Voting score | Score % 4 overall |
| Segment | Severity range | | |
| 12 | 0.00155~0.00165 | 0.4791 | 11.978% |
| 8 | 0.00135~0.00145 | 0.2316 | 5.790% |
| 21 | 0.00145~0.00155 | 0.1962 | 4.910% |
| 12 | 0.00145~0.00155 | 0.1919 | 4.798% |
| 22 | 0.00155~0.00165 | 0.1839 | 4.598% |

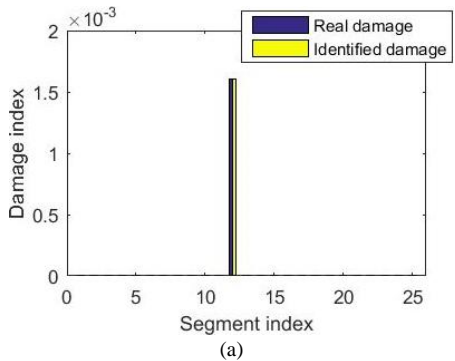

(a)

Finally, the four different means of handling voting scores used in case studies are compared in Figure 20. The voting power of the best solution in each case study is quantified in the form of voting score percentage, which is the percentage of its voting score out of available voting scores being distributed. As illustrated, by either grouping the damage severities to severity intervals or assigning voting scores to only certain solution sets, the confidence of the implications made by the proposed approach, which is directly related to the voting power, is increased. In practice, as we want to inspect only a small number of damage scenarios in maintenance, the overall approach proposed in this study can help to isolate a small set of the solutions that are more related to the health condition of the structure through its data-assisted analysis. Instead of seeking for one deterministic solution that could be misguiding, the approach proposed in this research utilizes training data to analyze and identify probable fault scenarios that serve as guidance for further examination through heterogeneous sensing and inspection.

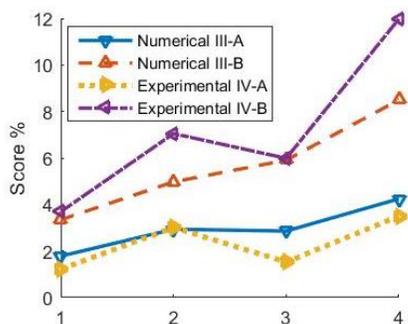

Figure 20 Voting score percentage of the best solution in each case study as four post-processing techniques adopted (Ticks on the horizontal axis represent: voting scores for damage scenarios, voting scores for damage severity ranges, partial voting scores for damage scenarios and partial voting scores for damage severity ranges, respectively from left to right)

## V. CONCLUDING REMARKS

This research presents a data-assisted approach for structural fault identification through Gaussian process-based impedance response calibration and many-objective evaluation. To address the fundamental challenges posed by the under-determined problem formulation and model-based sensitivity approximation, we cast the damage identification problem into a many-objective optimization by reconstructing impedance response surfaces as objective functions utilizing training data. The optimization problem is then tackled by an $\varepsilon$-dominance enabled many-objective simulated annealing algorithm. As many solutions are expected in many-optimization practices, a voting score calculation procedure is developed and applied after to quantify and identify the solutions that could make better implication about the health condition of the structure. The numerical case studies and experimental case studies demonstrate that the proposed approach is capable of obtaining a small set of solutions based on their voting scores that could provide accurate implication about the health condition of the interested structure. The proposed scheme is inherently malleable and can be applied to either model-based or model-free fault identification systems wherever data is available. The combination of Gaussian process-based calibration, many-objective optimization, and voting score calculation can be extended to a variety of inverse analysis problems.